\documentclass[prc,letterpaper,twocolumn,showpacs,showkeys,lengthcheck,tightenlines,
               nofootinbib,preprintnumbers,superscriptaddress]{revtex4-1} 

\pdfpageattr {/Group << /S /Transparency /I true /CS /DeviceRGB>>}
\usepackage[utf8]{inputenc}
\usepackage{newtxtext,newtxmath}

\usepackage{titlesec}
\usepackage{dcolumn}
\usepackage{bm}
\usepackage{amsmath,amssymb,amsfonts}

\usepackage{graphicx}

\usepackage{color}
\usepackage{bbold}

\usepackage{slashed}
\usepackage[svgnames]{xcolor}
\usepackage[english]{babel}
\usepackage{blindtext}
\usepackage{microtype}
\usepackage{tikz}

\usepackage[colorlinks=true,linkcolor=blue,urlcolor=blue,citecolor=blue]{hyperref}
\usepackage{ifpdf}

\def\L{\Lambda}

\def\hs{\hspace}

\def\no{\nonumber}

\def\lf{\left}
\def\rg{\right}

\newcommand{\vect}[1]{\boldsymbol{#1}}
\newcommand{\ph}[1]{\phantom{#1}}

\newcommand{\sh}[1]{\slashed{#1}}

\graphicspath{{.}{figures/}} 

\titlespacing{\section}{4pt}{10pt plus 4pt minus 2pt}{8pt plus 2pt minus 2pt}
\titlespacing{\subsection}{0pt}{12pt plus 4pt minus 2pt}{8pt plus 2pt minus 2pt}
\usepackage[mathscr,scaled=1.15]{urwchancal}
\DeclareFontFamily{OT1}{pzc}{}
\DeclareFontShape{OT1}{pzc}{m}{it}%
{<-> s * [1.15] pzcmi7t}{}
\DeclareMathAlphabet{\mathpzc}{OT1}{pzc}{m}{it}

\makeatletter
\newcommand*\if@single[3]{%
  \setbox0\hbox{${\mathaccent"0362{#1}}^H$}%
  \setbox2\hbox{${\mathaccent"0362{\kern0pt#1}}^H$}%
  \ifdim\ht0=\ht2 #3\else #2\fi
  }
\newcommand*\rel@kern[1]{\kern#1\dimexpr\macc@kerna}
\newcommand*\widebar[1]{\@ifnextchar^{{\wide@bar{#1}{0}}}{\wide@bar{#1}{1}}}
\newcommand*\wide@bar[2]{\if@single{#1}{\wide@bar@{#1}{#2}{1}}{\wide@bar@{#1}{#2}{2}}}
\newcommand*\wide@bar@[3]{%
  \begingroup
  \def\mathaccent##1##2{%
    \if#32 \let\macc@nucleus\first@char \fi
    \setbox\z@\hbox{$\macc@style{\macc@nucleus}_{}$}%
    \setbox\tw@\hbox{$\macc@style{\macc@nucleus}{}_{}$}%
    \dimen@\wd\tw@
    \advance\dimen@-\wd\z@
    \divide\dimen@ 3
    \@tempdima\wd\tw@
    \advance\@tempdima-\scriptspace
    \divide\@tempdima 10
    \advance\dimen@-\@tempdima
    \ifdim\dimen@>\z@ \dimen@0pt\fi
    \rel@kern{0.6}\kern-\dimen@
    \if#31
      \overline{\rel@kern{-0.6}\kern\dimen@\macc@nucleus\rel@kern{0.4}\kern\dimen@}%
      \advance\dimen@0.4\dimexpr\macc@kerna
      \let\final@kern#2%
      \ifdim\dimen@<\z@ \let\final@kern1\fi
      \if\final@kern1 \kern-\dimen@\fi
    \else
      \overline{\rel@kern{-0.6}\kern\dimen@#1}%
    \fi
  }%
  \macc@depth\@ne
  \let\math@bgroup\@empty \let\math@egroup\macc@set@skewchar
  \mathsurround\z@ \frozen@everymath{\mathgroup\macc@group\relax}%
  \macc@set@skewchar\relax
  \let\mathaccentV\macc@nested@a
  \if#31
    \macc@nested@a\relax111{#1}%
  \else
    \def\gobble@till@marker##1\endmarker{}%
    \futurelet\first@char\gobble@till@marker#1\endmarker
    \ifcat\noexpand\first@char A\else
      \def\first@char{}%
    \fi
    \macc@nested@a\relax111{\first@char}%
  \fi
  \endgroup
}
\makeatother

\begin{document}

\title{Intrinsic Transverse Motion of the Pion's Valence Quarks \vspace*{-0.3em}}

\author{Chao Shi}
\affiliation{Physics Division, Argonne National Laboratory, Argonne, IL 60439 USA}

\author{Ian C. Clo\"et}
\affiliation{Physics Division, Argonne National Laboratory, Argonne, IL 60439 USA}

\begin{abstract}
Starting with the solution to the Bethe-Salpeter equation for the pion, in a beyond rainbow-ladder truncation to QCD's Dyson-Schwinger equations (DSEs), we determine the pion's $l_z=0$ and $|l_z|=1$ leading Fock-state light-front wave functions (LFWFs) [labeled by $\psi_{l_z}(x,\vect{k}_T^2)$]. The leading-twist time-reversal even transverse momentum dependent parton distribution function (TMD) of the pion is then directly obtained from these LFWFs. A key characteristic of the LFWFs, which is driven by dynamical chiral symmetry breaking, is that at typical hadronic scales they are broad functions in the light-cone momentum fraction $x$. The LFWFs have a non-trivial $(x,\vect{k}_T^2)$ dependence and in general do not factorize into separate functions of each variable. The $l_z=0$ LFWF is concave with a maximum at $x=1/2$, whereas orbital angular momentum effects causes the $|l_z|=1$ LFWF to have a slight {\it double-humped} structure for quark transverse momentum in the range $0.5 \lesssim \vect{k}_T^2 \lesssim 5\,$GeV$^2$. For $\vect{k}_T^2 \lesssim 1\,$GeV$^2$ the $\vect{k}_T^2$ dependence of the LFWFs is well described by a Gaussian, however for $\vect{k}_T^2 \gtrsim 10\,$GeV$^2$ these LFWFs behave as $\psi_0 \propto x(1-x)/\vect{k}_T^2$  and  $\psi_1 \propto x(1-x)/\vect{k}_T^4$, and therefore exhibit the power-law behavior predicted by perturbative QCD. The pion's TMD inherits many features from the LFWFs, where for $\vect{k}_T^2 \lesssim 1\,$GeV$^2$ the $\vect{k}_T^2$ dependence is well described by a Gaussian, and for large $\vect{k}_T^2$ the TMD behaves as $f_\pi^q \propto x^2(1-x)^2/\vect{k}_T^4$. At the model scale we find the average transverse momentum, defined by a Bessel-weighted moment with $\vect{b}_T=0.3\,$fm, to equal $\big<\vect{k}_T^2\big> = 0.19\,$GeV$^2$. The TMD evolution of our result is studied using both the $b^*$ and $\zeta$ prescriptions which allows a qualitative comparison with existing Drell-Yan data.
\end{abstract}

%\noindent PACS Number(s): 12.38.Aw, 12.38.Lg, 11.10.St,  14.40.Be

\maketitle
%===============================================================================
%===============================================================================
Light-front quantization and the associated light-front wave functions (LFWFs) provide a powerful framework with which to study quantum chromodynamics (QCD) and develop an understanding of the parton structure of hadrons~\cite{Brodsky:1997de,Heinzl:2000ht}. Hadron observables such as form factors, parton distribution functions (PDFs), and their multi-dimensional counterparts such as generalized and transverse momentum dependent PDFs (TMDs) can each be expressed as overlaps of LFWFs~\cite{Brodsky:2000xy,Pasquini:2014ppa}. Therefore LFWFs allow features of apparent disparate hadron observables to be straightforwardly related to underlying quark-gluon dynamics in a QCD Fock-state expansion. In principle, the LFWFs can be computed by diagonalizing the light-front QCD Hamiltonian operator, using methods such as discretized light-cone quantization~\cite{Pauli:1985ps}, or basis light-front quantization~\cite{Vary:2009gt,Li:2017mlw}. However, these calculations become numerically challenging for QCD in four space-time dimensions, therefore effective interactions such as holographic QCD have been used to reduce these difficulties~\cite{Brodsky:2014yha}.

Another approach used to study QCD and hadron structure, which is explicitly Poincar\'e-covariant, is provided by judicious truncations to QCD's Dyson-Schwinger equations (DSEs)~\cite{Roberts:1994dr,Alkofer:2000wg,Cloet:2013jya}. In the DSE framework hadron states are obtained as solutions to Poincar\'e-covariant bound-state equations such as the Bethe-Salpeter and Faddeev equations~\cite{Cloet:2008re,Eichmann:2009qa}. Insights into numerous aspects of hadron structure have been revealed using the DSEs~\cite{Cloet:2013jya,Eichmann:2016yit}, with particular success in understanding the pion as both a relativistic bound-state of a dressed quark and dressed antiquark, and the Goldstone mode associated with dynamical chiral symmetry breaking (DCSB) in QCD~\cite{Maris:1997hd,Chang:2013nia,Chang:2013pq,Cloet:2013jya}. DSE solutions to the Bethe-Salpeter equation (BSE), which encapsulate key emergent QCD phenomena such as DCSB and quark confinement, therefore provide an excellent starting point from which to extract the pion's LFWFs. In particular, the properties of the LFWFs can then be clearly connected to underlying quark-gluon dynamics as expressed in the dressing functions for propagators and vertices. The calculation of the pion's leading Fock-state LFWFs using the DSEs, and the application of these LFWFs to a calculation of the pion's leading-twist time-reversal even TMD is the main focus of this paper. Such a study is timely because the proposed electron-ion collider~\cite{longrangeplan} has the capability to study the partonic structure of the pion and kaon~\cite{pieic:2018}.

In the light-front formalism a hadron state can be expressed as the superposition of Fock-state components classified by their orbital angular momentum projection $l_z$~\cite{Burkardt:2002uc}. For the pion the  minimal ($\bar{q}q$) Fock-state configuration reads~\cite{Burkardt:2002uc,Ji:2003yj}: 
\begin{align}
\lf|\pi^+(p) \rg> &=|\pi^+(p)\rangle_{l_z=0}+|\pi^+(p)\rangle_{|l_z|=1}, 
\label{eq:pionLFWF}
\end{align}
where the non-perturbative content of each state is contained in the LFWFs~\cite{Pasquini:2014ppa}, labeled by $\psi_0(x,\vect{k}_T^2)$ for $l_z=0$ and $\psi_1(x,\vect{k}_T^2)$ for $|l_z|=1$, where $\vect{k}_T$ is the transverse momentum of the quark and $x=\frac{k^+}{p^+}$ is its light-cone momentum fraction. For these minimal Fock-state LFWFs the antiquark has transverse momentum $-\vect{k}_T$ (in a frame where $\vect{p}_T = 0$ for the pion) and light-cone momentum fraction $1-x$.

From the matrix element definitions of the LFWFs~\cite{Burkardt:2002uc}, it can be shown that the pion's minimal Fock-state LFWFs can be obtained from the pion's Poincar\'e-covariant Bethe-Salpeter wave function, $\chi(k,p)$, via~\cite{Mezrag:2016hnp}
\begin{align}
\label{eq:psi0}
\psi_0(x,\vect{k}_T^2)&= \ph{-}\sqrt{3}\,i\!\int \frac{dk^+dk^-}{2\,\pi} \no \\
& \hs*{11mm} \times
\textrm{Tr}_D\!\left[ \gamma^+ \gamma_5 \chi(k,p)\right]  \delta\lf(x\,p^+ -k^+\rg),  \\
\label{eq:psi1}
\psi_1(x,\vect{k}_T^2)&= -\sqrt{3}\,i\!\int \frac{dk^+dk^-}{2\,\pi}\,  \frac{1}{\vect{k}_T^2} \no \\ 
& \hs*{3mm} \times
\textrm{Tr}_D\left[ i\sigma_{+ i} \vect{k}_{T}^i\, \gamma_5\, \chi(k,p) \right] 
 \delta\lf(x\,p^+ -k^+\rg),   
\end{align}
where the trace is over Dirac indices only. The Bethe-Salpeter wave function for the $\pi^+$ is defined by the quark-antiquark correlator $\chi(k,p) = \int d^4 z\ e^{-ik \cdot z}\, \langle 0|\mathcal{T} u(z)\,\bar{d}(0)| \pi^+(p)\rangle$~\cite{Itzykson:1980rh,Gromes:1992ph} and can be expressed as $\chi(k,p) = S(k)\,\Gamma(k,p)\,S(k-p)$, where $S(k)$ is the dressed quark propagator and $\Gamma(k,p)$ the pion's homogeneous Bethe-Salpeter amplitude~\cite{LlewellynSmith:1969az,Roberts:1994dr}. 

The pion's Bethe-Salpeter wave function can be calculated within the DSE framework~\cite{Maris:2003vk}. This is achieved via a self-consistent solution to the quark gap equation which gives the dressed quark propagator $S(k)$, and the homogeneous BSE which gives the Bethe-Salpeter vertex $\Gamma(k,p)$. To obtain a solution to these equations we must employ a truncation to the interaction kernel such that the key symmetries of QCD are maintained. In the context of the pion the axial-vector Ward-Takahashi identity (WTI) plays an important role~\cite{Maris:1997tm}, as it is an expression of chiral symmetry and its dynamical breaking~\cite{Chang:2011ei}. The simplest symmetry-preserving DSE truncation is known as rainbow-ladder and has achieved many successes in the study of hadron properties~\cite{Maris:1997tm,Maris:1999nt,Maris:2000sk}. In this work we utilize a modern extension known as the DCSB-improved truncation. A key feature of this truncation to the DSEs is the presence of an anomalous chromomagnetic moment term in the dressed quark-gluon vertex~\cite{Chang:2010hb}, which in the chiral limit can only exist through DCSB. This truncation provides the most realistic description of the pion, and other hadrons, currently available within DSEs formalism~\cite{Cloet:2013jya}. 

The DSEs are formulated in Euclidean space and therefore a direct calculation of light-cone dominated quantities is not straightforward. However, an arbitrary $\vect{k}_T^2$-dependent moment of the pion's LFWFs, defined by
\begin{align}
\lf< x^m \rg>_{l_z}(\vect{k}^2_T) &= \int_0^1 dx\, x^m\, \psi_{l_z}(x,\vect{k}_T^2),
\label{eq:momsM}
\end{align}
can be directly calculated using the DSEs, and the LFWFs for the pion can then be accurately reconstructed from these moments. In fact, it will prove possible to express an arbitrary moment of a LFWF in the form~\cite{Cloet:2005pp,Mezrag:2016hnp}
\begin{align}
\label{eq:ced}
\lf< x^m \rg>_{l_z}(\vect{k}^2_T) = \int_0^1 d\alpha\, \alpha^m \int d\beta d\gamma\,  f_{l_z}(\alpha,\vect{k}_T^2,\beta,\gamma),
\end{align}
and therefore the LFWF is identified as $\psi_{l_z}(x,\vect{k}_T^2)= \int d\beta d\gamma \, f_{l_z}(x,\vect{k}_T^2,\beta,\gamma)$. 

To aid the calculation of the moments $\lf< x^m \rg>_{l_z}$ we use an accurate parametrization of numerical solutions to the gap and BSEs in the DCSB-improved truncation to the DSEs~\cite{Chang:2010hb,Chang:2013pq}. The dressed quark propagator is parametrized with two pairs of complex conjugate poles~\cite{Tiburzi:2003ja,Souchlas:2010boa}: $S(k)=\sum_{i=1}^{2} [z_i/(i\sh{k} +m_i) + z^*_i/(i\sh{k} + m^*_i)]$ where $z_i$ and $m_i$ are complex numbers determined by fitting to the numerical DSE solution to the gap equation. The general Bethe-Salpeter amplitude for the pion reads~\cite{LlewellynSmith:1969az,Maris:1997hd}: $\Gamma_\pi(k,p) = \gamma_5\big[iE(k,p) + \sh{p}\,F(k,p) + \sh{k}\, G(k,p) + [\sh{p},\sh{q}]\,H(k,p)\big]$. Herein we retain the dominant $E$ and $F$ amplitudes, which are accurately parameterized by ($f = E, F$)
\begin{align}
\label{eq:fpara}
f(k,p) &= \mathcal{A}^f_{\rm IR}(k,p) + \mathcal{A}^f_{\rm UV}(k,p), \allowdisplaybreaks \\
\mathcal{A}^i_{\rm IR}(k,p) &= a_1^i \int_{-1}^1 dz\ \rho(z;\sigma_1^i) \no \\
&\hs*{6mm}
\lf[b^i\,\Delta^4(k_z^2;\Lambda_1^i) + \lf(1 - b^i\rg)\,\Delta^5(k_z^2;\Lambda_1^i) \rg], \\
\mathcal{A}^i_{\rm UV}(k,p) &= a_2^i \int_{-1}^1dz\ \rho(z;\sigma_2^i)\ \Delta(k_z^2;\Lambda_2^i),
\label{eq:B}
\end{align}
with $k_z^2 \equiv k^2 + z\,k \cdot p$, $\Delta(s;\Lambda) = 1/[1 + s/\Lambda^2]$,
$\rho(z;\sigma) = \frac{1}{2}\lf[C_0^{(1/2)}(z) + \sigma\,C_2^{(1/2)}(z)\rg]$, 
and $C_n^{(\alpha)}$ are Gegenbauer polynomials. This work introduces a new form for the weight function $\rho$ which gives a more faithful representation of observables that are sensitive to the LFWFs near $x=0,1$. The fit parameters are given in Tab.~\ref{Table:parameters}.

%==============================================================================
\begin{table}[tbp]
\addtolength{\extrarowheight}{2.2pt}
\begin{tabular*}%{lcccc}
{\hsize}
{
c@{\extracolsep{0ptplus1fil}}
c@{\extracolsep{0ptplus1fil}}
c@{\extracolsep{0ptplus1fil}}
c@{\extracolsep{0ptplus1fil}}}
\hline\hline
$z_1$ & $m_1$  & $z_2$ & $m_2$ \\
\hline
$(0.44,0.28)$ & $(0.46,0.18)$ & $(0.12,0.00)$ & $(-1.31,-0.75)~~$ \\
\hline\hline\\[-0.5em]
\end{tabular*}
\begin{tabular*}%{llcccccccc}
{\hsize}
{
l@{\extracolsep{0ptplus1fil}}
c@{\extracolsep{0ptplus1fil}}
c@{\extracolsep{0ptplus1fil}}
c@{\extracolsep{0ptplus1fil}}
c@{\extracolsep{0ptplus1fil}}
c@{\extracolsep{0ptplus1fil}}
c@{\extracolsep{0ptplus1fil}}
c@{\extracolsep{0ptplus1fil}}}
\hline\hline
& $a_1$ & $a_2$ & $b$ & $\sigma_1$ & $\sigma_2$ & $\Lambda_1$ & $\Lambda_2~~$\\
\hline
$E$ & $0.92$ & $3.0$ & $0.08\ph{0}$  & $\ph{-}2.2$ & $-1.0$ & $1.41$ & $1.0~~$\\
$F$ & $0.55$ & $3.0$ & $0.008$       & $-0.5$      & $-1.0$ & $1.13$ & $1.0~~$  \\\hline\hline
\end{tabular*}
\caption{\emph{Upper panel}: Complex conjugate pole parameters that give an accurate representation of the DCSB-improved quark propagator.
\emph{Lower panel}: Parameters used in Eqs.~\eqref{eq:fpara}--\eqref{eq:B} to represent the DCSB-improved solution for the pion's homogeneous Bethe-Salpeter vertex. In both panels all dimensioned quantities are in units of GeV.}
\label{Table:parameters}
\end{table}
%==============================================================================

\begin{figure}[tbp]
\centering\includegraphics[width=\columnwidth]{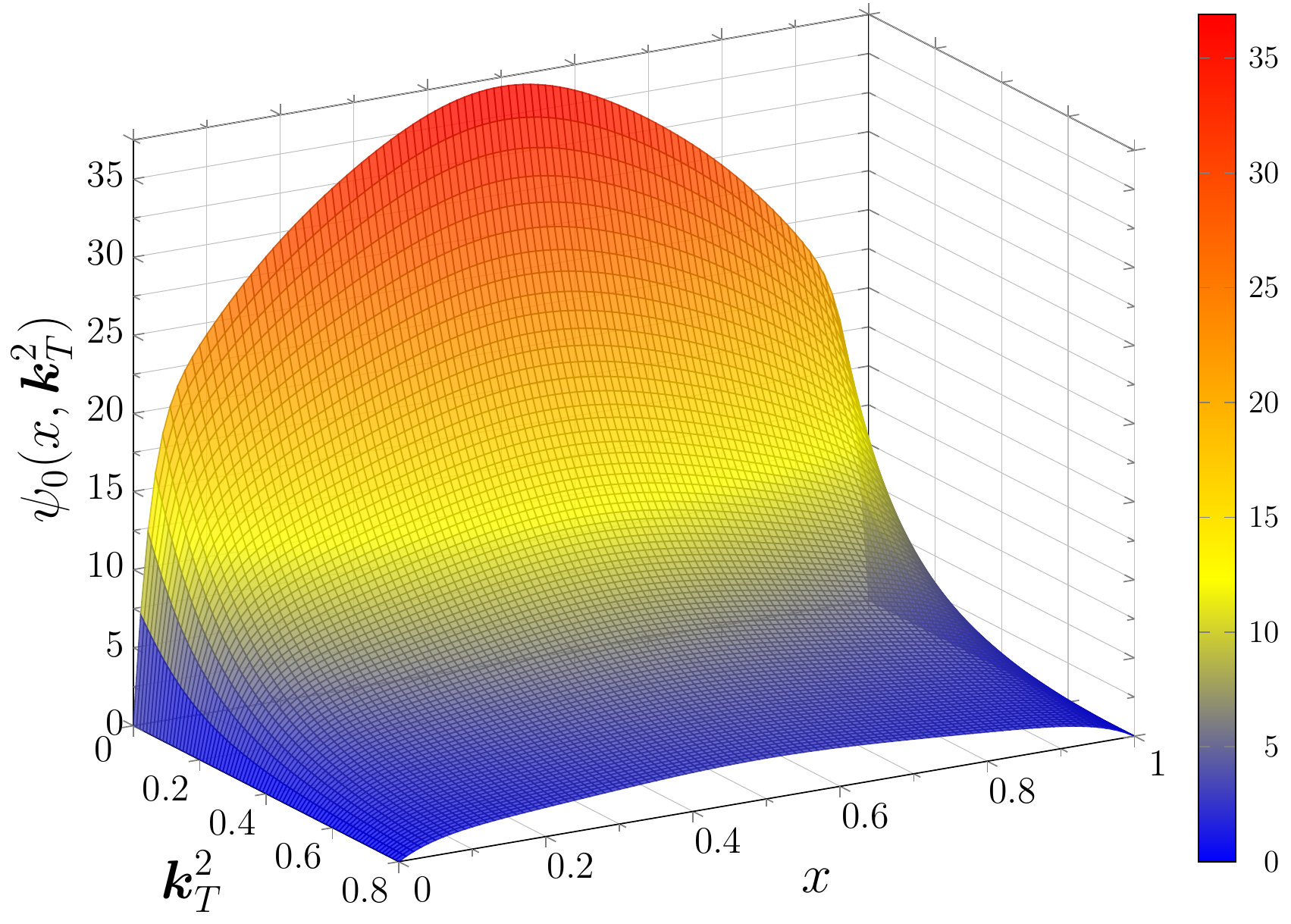} \\
\centering\includegraphics[width=\columnwidth]{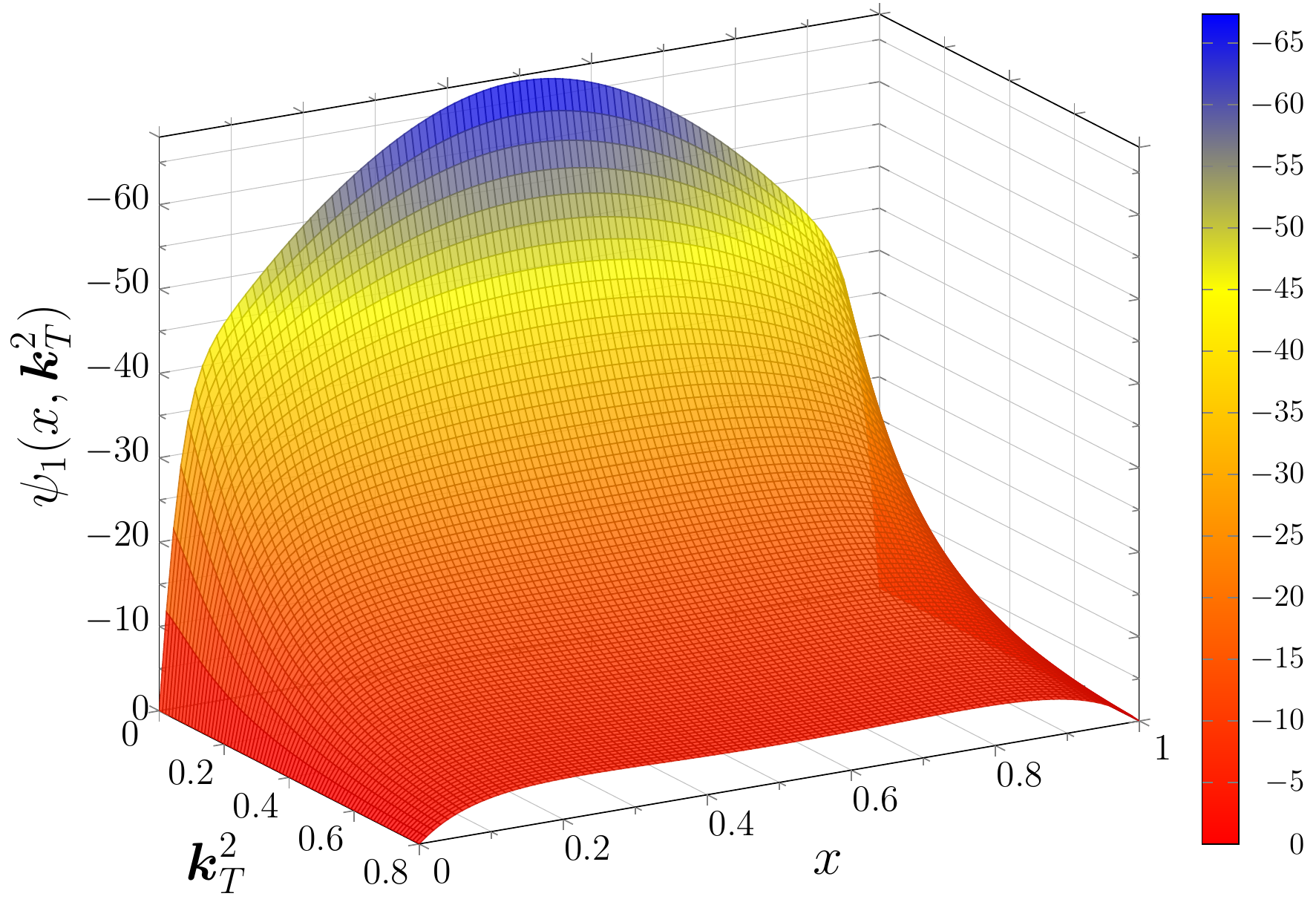}
\caption{\looseness=-1
{\it Upper panel:} DSE result using the DCSB-improved kernel for the pion's $l_z=0$  minimal ($\bar{q}q$) Fock-state LFWF. 
{\it Lower panel:} Analogous result for the pion's $|l_z|=1$ minimal Fock-state LFWF.
The LFWFs are given in units of GeV$^{-2}$ and $\vect{k}^2_T$ in GeV$^2$.
}
\label{fig:psi0psi1}
\end{figure}

Our results for the pion's minimal ($\bar{q}q$) Fock-state LFWFs are illustrated in Fig.~\ref{fig:psi0psi1}, where the LFWFs satisfy the normalization condition $\int_0^1\! dx\! \int\! \frac{d^2 \vect{k}_T}{(2\pi)^3} \Big[|\psi_0(x,\vect{k}_T^2)|^2 + \vect{k}_T^2|\psi_1(x,\vect{k}_T^2)|^2 \Big] = 1$. For each $x$, the $\vect{k}_T^2$ dependence of the LFWFs exhibits a Gaussian-like behavior for $\vect{k}_T^2 \lesssim 1\,$GeV$^2$, a transition then begins to occur and for $\vect{k}_T^2 \gtrsim 10\,$GeV$^2$ the LFWFs become $\psi_0(x,\vect{k}_T^2) \propto x(1-x)/\vect{k}_T^2$ and  $\psi_1(x,\vect{k}_T^2) \propto x(1-x)/\vect{k}_T^4$, which matches the power-law behavior predicted by perturbative QCD~\cite{Ji:2003yj}. We therefore predict that the $x$-dependence of 
$\psi_0$ and $\psi_1$ is the same for large $\vect{k}_T^2$. A factorization between $x$ and $\vect{k}_T^2$ is only seen in the scaling regime, where the onset reflects the ultraviolet behavior of the Bethe-Salpeter dressing functions which behave as $E, F \sim 1/k^2$  for $k^2 \gtrsim 10\,$GeV$^2$~\cite{Maris:1997tm}. 

An important characteristic of our LFWF results, when viewed as a function of $x$, is that they are broad  with significant support near the $x=0,\,1$ end-points for $\vect{k}^2_T \lesssim 1\,$GeV$^2$. As discussed in Ref.~\cite{Chang:2013pq} in the context of the pion's parton distribution amplitude (PDA), this broadening of the LFWFs is directly linked to DCSB, however this effect diminishes for $\vect{k}^2_T \gg \Lambda_{\rm QCD}^2$ where we find that the $x$-dependence of the LFWFs is the same as the asymptotic pion PDA~\cite{Lepage:1980fj}. This manifestation of DCSB on the light-front will therefore have a material impact on observables sensitive to the region $\vect{k}^2_T \lesssim 1\,$GeV$^2$ for all $x$. The $l_z=0$ LFWF is concave with a maximum at $x=1/2$ for all $\vect{k}_T^2$, whereas orbital angular momentum effects causes the $|l_z|=1$ LFWF to have a slight {\it double-humped} structure for quark transverse momentum in the range $0.5 \lesssim \vect{k}_T^2 \lesssim 5\,$GeV$^2$, which is evident in Fig.~\ref{fig:psi0psi1}.  Near the $x=0,\,1$ end-points we find that each LFWF behaves linearly as a function of $x$, that is, as $x\rightarrow 1$ we have $\psi_{l_z}(x,\vect{k}_T^2) \sim 1-x$, with analogous results near $x\to 0$ because $\psi_{l_z}(x,\vect{k}_T^2) = \psi_{l_z}(1-x,\vect{k}_T^2)$. This linear behavior in $1-x$ is a necessary property of the LFWFs if they are to give a pion TMD or PDF behaving as $f(x) \to (1-x)^2$ near $x=1$, as predicted by perturbative QCD~\cite{Farrar:1975yb,Ji:2004hz,Brodsky:2006hj}.

%===============================================================================
%===============================================================================

With the pion's LFWFs in hand it is now straightforward to determine numerous properties of the pion. Here we focus on the pion's leading-twist time-reversal even TMD which in terms of the pion's minimal Fock-state LFWFs reads~\cite{Pasquini:2014ppa}
\begin{align}
\hs*{-1.5mm}f_{\pi}^{\mu_0}(x,\vect{k}_T^2) = 
%\frac{1}{(2 \pi)^3} \left[|\psi_0^{\mu_0}(x,\vect{k}_T^2)|^2 + \vect{k}_T^2 |\psi_1^{\mu_0}(x,\vect{k}_T^2)|^2\right], 
\left[|\psi_0^{\mu_0}(x,\vect{k}_T^2)|^2 + \vect{k}_T^2 |\psi_1^{\mu_0}(x,\vect{k}_T^2)|^2\right]/(2 \pi)^3, 
\label{eq:tmd}
\end{align}
where we have made explicit the renormalization scale dependence of the LFWFs and consequently the TMD. The pion's valence quark PDF is related to the TMD by $f_{\pi}^{\mu_0}(x) = \int d^2\vect{k}_T\ f_{\pi}^{\mu_0}(x,\vect{k}_T^2)$, where the normalization condition for the LFWFs guarantees baryon number conservation ($\langle x^0 \rangle^{\mu_0} = 1$). The symmetry under $x \to 1-x$ of the LFWFs also ensures $\langle x \rangle^{\mu_0} = 0.5$ and therefore the two valence quarks carry all the momentum of the pion. If one associates the renormalization scale with the resolving scale ($\mu_0^2 = Q^2$), then as $\mu_0^2$ gets larger higher Fock-state amplitudes begin to play an increasingly important role, and therefore the minimal Fock-state contributes calculated here are only dominant at a low resolving scale~\cite{Burkardt:2002uc}. 
%
%\textcolor{blue}{A}ssuming that at a low hadronic scale $\mu_0$, the pion is well approximated by a dressed quark-antiquark pair where all sea-quarks and gluons can be absorbed into these degrees of freedom~\cite{Bacchetta:2017vzh}. 
The renormalization scale associated with our DSE calculation is determined such that the momentum fraction carried by the valence quarks agrees with results from a $\pi N$ Drell-Yan analysis $2\,\langle x \rangle_v = 0.47(2)$~\cite{Sutton:1991ay,Gluck:1999xe} or the lattice QCD result $2\,\langle x \rangle_v = 0.48(4)$~\cite{Detmold:2003tm} both at a scale of $Q^2 = 4\,$GeV$^2$. Using NLO DGLAP~\cite{Botje:2010ay} we obtain a model scale of $\mu_0 = 0.52$\,GeV. 

%===============================================================================
\begin{figure}[tbp]
\centering\includegraphics[width=\columnwidth]{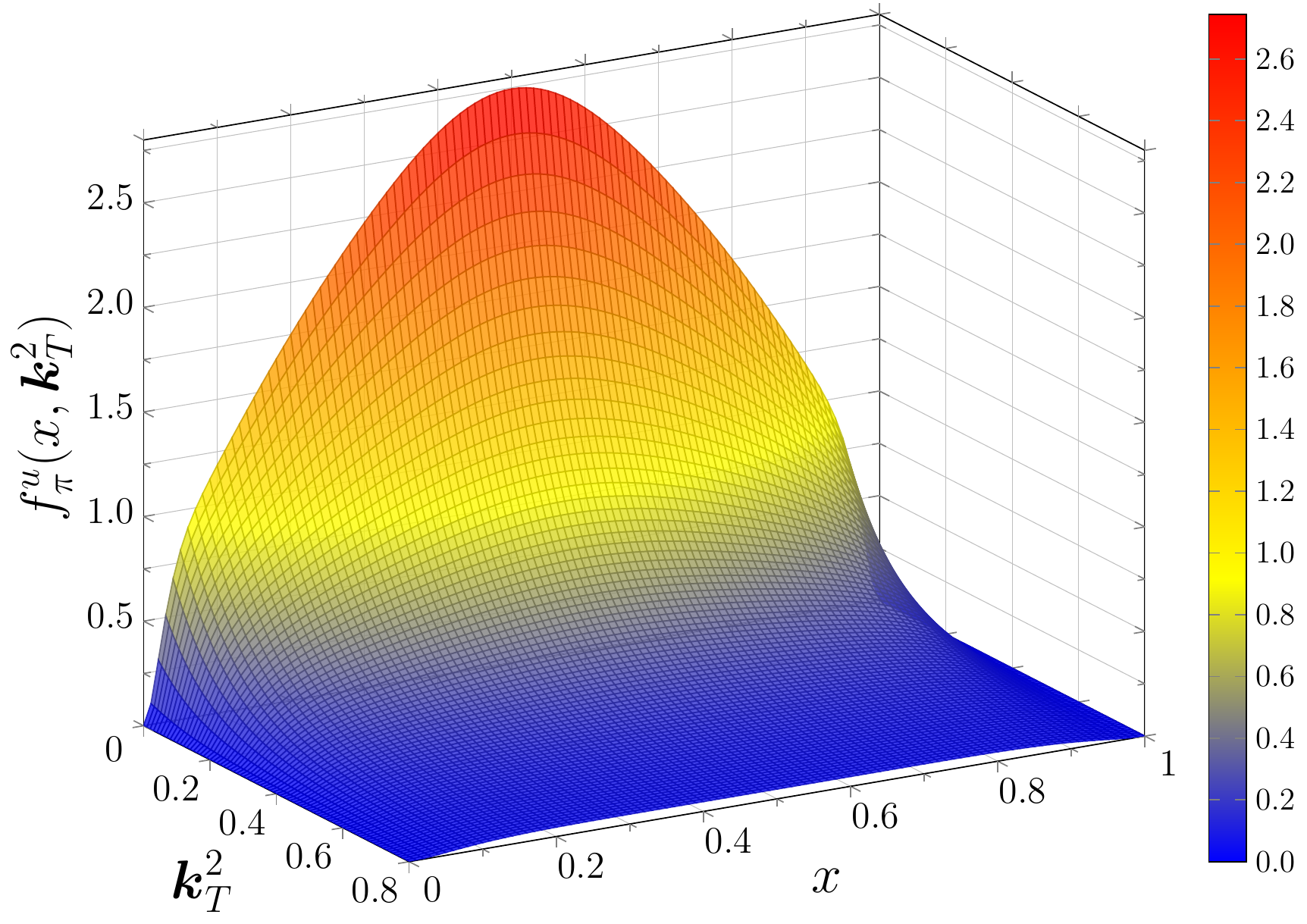} \\
\centering\includegraphics[width=\columnwidth]{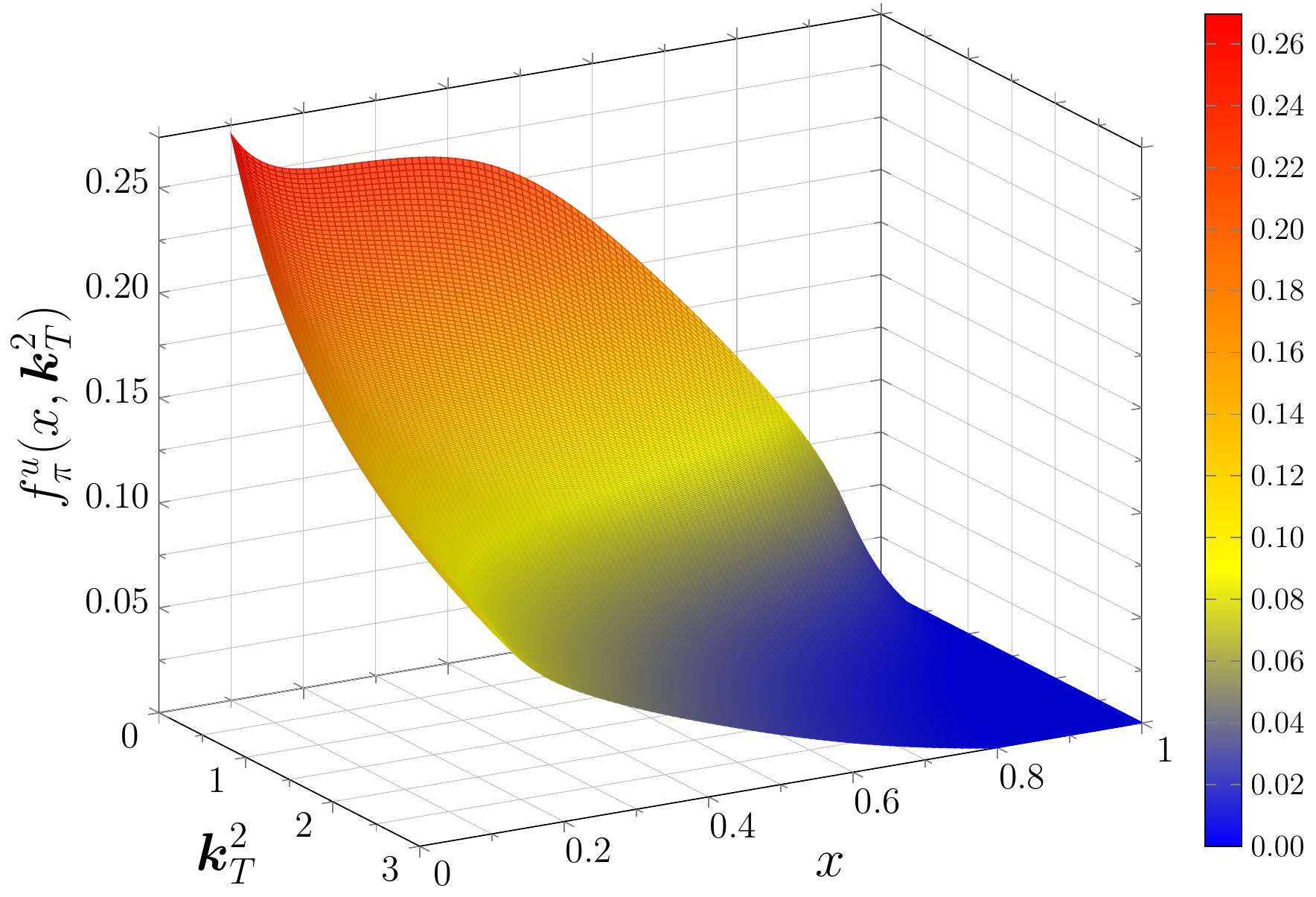}
\caption{
{\it Upper panel:} DSE result using the DCSB-improved kernel for the time-reversal even $u$-quark TMD of the pion, $f_\pi^u(x,\vect{k}_T^2)$, at the model scale of $\mu^2_0 = 0.52\,$GeV$^2$.
{\it Lower panel:} Analogous result evolved to a scale of $\mu = 6\,$GeV using TMD evolution with the $b^*$ prescription and $g_2 = 0.09\,$GeV~\cite{Bacchetta:2017gcc}. The TMDs are given in units of GeV$^{-2}$ and $\vect{k}^2_T$ in GeV$^2$.} 
\label{fig:tmds}
\end{figure}
%===============================================================================

Our DSE result for the time-reversal even $u$-quark TMD in the $\pi^+$, obtained from the LFWFs using Eq.~\eqref{eq:tmd}, is given in the upper panel of Fig.~\ref{fig:tmds}. These calculations are performed with equal current quark masses ($m_u = m_d$), and therefore the $\bar{d}$ TMD in the $\pi^+$ is the same as the $u$-quark TMD. Several features of the LFWFs are immediately reflected in the TMD at the hadronic scale, notably in the limit $x\to 1$ the TMD behaves as $f_{\pi}^u(x,\vect{k}_T^2) \propto (1-x)^2$ for all $\vect{k}^2_T$, in agreement with perturbative QCD~\cite{Brodsky:2006hj}. As $\vect{k}^2_T$ becomes large our TMDs exhibits two scaling regimes, for $\vect{k}_T^2 \gtrsim 10\,$GeV$^2$ the pion's TMD has a power-law behavior of $f_{\pi}^u(x,\vect{k}_T^2) \propto 1/\vect{k}_T^6$ which reflects the dominance of $\psi_1(x,\vect{k}_T^2)$ in this region. The $l_z=0$ LFWF only begins to dominate the TMD for $\vect{k}_T^2 \gtrsim 100\,$GeV$^2$, where we obtain our asymptotic result for the TMD: $f_\pi^u(x,\vect{k}_T^2) \propto x^2(1-x)^2/\vect{k}_T^4$. At the low hadron scale our DSE result for the pion's TMD is a broad unimodal function of $x$ for $\vect{k}_T^2 \lesssim 0.7\,$GeV$^2$, however in the range $0.7 \lesssim \vect{k}_T^2 \lesssim 5\,$GeV$^2$ the {\it double-humped} feature of $\psi_1(x,\vect{k}_T^2)$ manifests in the TMD. We stress that the double-humped structure we see in our result for the TMD is slight, as made clear from the upper panel in Fig.~\ref{fig:tmds}, however this structure is seen more prominently in some light-front constituent quark~\cite{Pasquini:2014ppa} and holographic QCD models~\cite{Bacchetta:2017vzh}. Because our TMD result scales as $f_\pi^u(x,\vect{k}_T^2) \propto 1/\vect{k}_T^4$, our result for the average $\vect{k}_T^2$ of the TMD is logarithmically divergent if $\big<\vect{k}_T^2\big>$ is defined in the usual way~\cite{Avakian:2010br}. We therefore study two methods: fitting a Gaussian ansatz to our TMD for $\vect{k}_T^2 < 1\,$GeV$^2$ gives $\big<\vect{k}_T^2\big> = 0.16\,$GeV$^2$, and using the Bessel-weighted definition proposed in Ref.~\cite{Boer:2011xd} with $\vect{b}_T = 0.3\,$fm gives $\big<\vect{k}_T^2\big> = 0.19\,$GeV$^2$ at the model scale. Therefore the scale of the average transverse momentum is typical of the infrared scale of the dressed quark mass, $M \simeq 400\,$MeV.

For a meaningful comparison between our results and (potential) data from {\it e.g.} semi-inclusive deep inelastic scattering and Drell-Yan experiments, it is essential to perform TMD evolution~\cite{Collins:2011zzd,Aybat:2011zv} of our model scale result. TMD evolution is governed by renormalization group equations involving two scales, $\mu$ and $\zeta$, which are set to the hard scale $\mu^2 = \zeta = Q^2$~\cite{Rogers:2015sqa}. The lower panel of Fig.~\ref{fig:tmds} presents our pion TMD result evolved to a scale of $\mu = 6\,$GeV, which is a typical scale associated with the E-615 pion-induced Drell-Yan experiment~\cite{Conway:1989fs}. The illustrated result uses the $b^*$-prescription~\cite{Collins:1981va}, where we follow closely the implementation of Ref.~\cite{Bacchetta:2017gcc,Bacchetta:2017vzh},  and to parameterize the non-perturbative behavior of the $\zeta$ evolution kernel~\cite{Aybat:2011zv} we choose $g_2 = 0.09$ in accordance with Ref.~\cite{Bacchetta:2017vzh}. The effect of the TMD evolution is dramatic, shifting significant strength to small $x$ and large $\vect{k}_T^2$, with a factor of 10 reduction in the magnitude of the TMD near $x \sim 1/2,~\vect{k}_T^2\sim 0$ compared to the model scale result. For the evolved TMD we find $\big<\vect{k}_T^2\big> = 0.69\,$GeV$^2$ using the Gaussian fit method, and the Bessel-weighted definition with $\vect{b}_T = 0.3\,$fm gives $\big<\vect{k}_T^2\big> = 0.49\,$GeV$^2$.\footnote{We also studied the TMD evolution of our result using the $\zeta$-prescription~\cite{Scimemi:2017etj} and find qualitatively similar results.}

To attempt a quantitative comparison of our results with data we study the transverse momentum dependence characterized by a fitting function $P(x_F,\vect{p}_T;m_{\mu \mu})$ measured in the E-615 pion-induced Drell-Yan experiment on a tungsten target~\cite{Conway:1989fs}. This function is defined by 
\begin{align}
\frac{d^3 \sigma}{dx_\pi d x_N d\vect{p}_T} = \frac{d^2 \sigma}{d x_\pi dx_N}\ P(x_F,\vect{p}_T;m_{\mu \mu}),
\end{align}
where $x_\pi,\,x_N$ are the Bjorken scaling variables of the pion and nucleon, $x_F = x_\pi - x_N$, and $m_{\mu \mu}^2 = s\,x_\pi x_N$ is the invariant mass-squared of the dilepton pair where $s = (p_\pi + p_N)^2$ is the center-of-mass energy squared. For the fitting function $P$ we have the relation $P(x_F,\vect{p}_T;m_{\mu \mu}) / |\vect{p}_T|\propto\,F^1_{UU}(x_\pi,x_N,\vect{p}_T)$, where within the TMD factorization scheme, at leading twist, the unpolarized Drell-Yan structure function is given by~\cite{Arnold:2008kf,Pasquini:2014ppa}
\begin{align}
F^1_{UU}(x_\pi,x_N,\vect{p}_T) &= \frac{1}{N_c}\sum_q e_q^2 \int d^2 \vect{k}_{T}d^2 \vect{\ell}_{T} \no \allowdisplaybreaks \\
&\hs*{-16mm} \times \delta^{(2)}(\vect{p}_T - \vect{k}_{T} - \vect{\ell}_{T})\,
f_{\pi}^{\bar{q}}(x_\pi, \vect{k}_{T}^2)\,f_A^q(x_N,\vect{\ell}_{T}^2),
\label{eq:fuu}
\end{align}
where the sum is over quark flavors $q = u,\,d$, and we approximate the unpolarized TMD of the tungsten target by a sum over nucleon TMDs: $f_A^q(x_N,\vect{\ell}_{T}^2) = Z/A\, f_p^q(x,\vect{\ell}_{T}^2) + N/A\, f_n^q(x,\vect{\ell}_{T}^2)$.  To evaluate $F^1_{UU}(x_\pi,x_N,\vect{p}_T)$ and thereby make a qualitative comparison with data for $P(x_F,\vect{p}_T;m_{\mu \mu})$ obtained in the E-615 experiment~\cite{Conway:1989fs} we combine our DSE results for $f_{\pi}^q(x_\pi, \vect{k}_{T}^2)$ with two sets of empirical extractions of $f_{p}^q(x,\vect{\ell}_{T}^2)$ and $f_{n}^q(x,\vect{\ell}_{T}^2)$ from Refs.~\cite{Bacchetta:2017gcc} and~\cite{Scimemi:2017etj} respectively.

%===============================================================================
\begin{figure}[tbp]
\centering\includegraphics[width=\columnwidth]{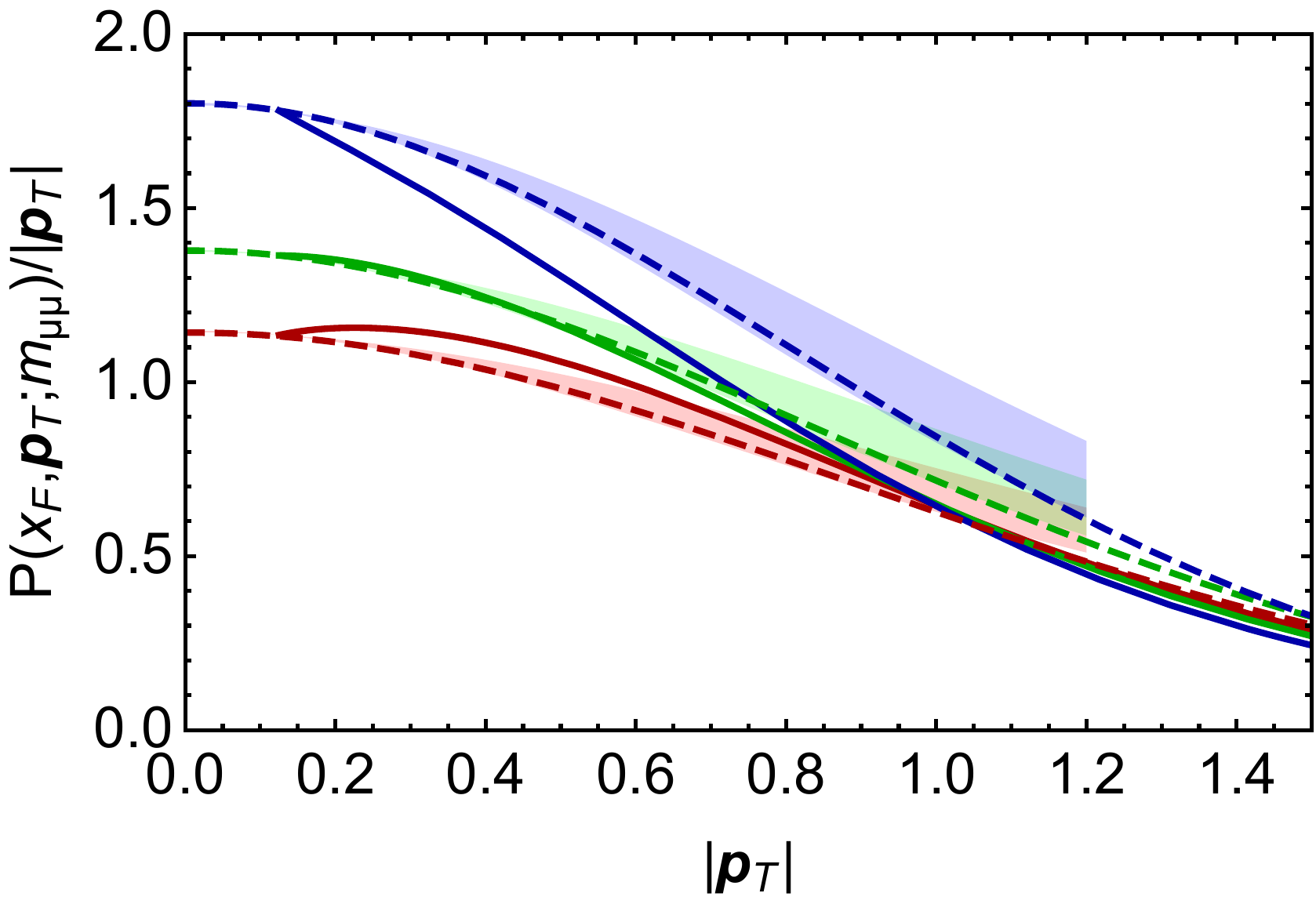}
\caption{\label{fig:Pfunction} The solid lines are empirical results from the E-615 experiment~\cite{Conway:1989fs} for the fitting function $P(x_F,\vect{p}_T;m_{\mu\mu})/|\vect{p}_T|$, the curves in ascending order correspond to $x_F = 0,\,0.25,\,0.5$. The neighboring shaded bands correspond to the same $x_F$ values, and are our results evolved using the $b^*$-prescription as outlined in Ref.~\cite{Bacchetta:2017gcc}, with the non-perturbative parameter $g_2$ in the range $0 \leqslant g_2 \leqslant 0.09$ (the lower boundary corresponds to $g_2=0$). The dashed lines are obtained using the $\zeta$-prescription from Ref.~\cite{Scimemi:2017etj} with $g_2=0$.}
\label{fig:p}
\end{figure}
%===============================================================================

Results for the fitting function $P(x_F,\vect{p}_T;m_{\mu \mu}) / |\vect{p}_T|$ are presented in Fig.~\ref{fig:p}. The solid lines are empirical results from Ref.~\cite{Conway:1989fs} for $x_F = 0,\,0.25,\,0.5$ where empirically $m_{\mu\mu} \simeq 6\,$GeV and $\sqrt{s} = 22\,$GeV. The shaded regions in Fig.~\ref{fig:p} are our calculated results for $\mathcal{N}\,F^1_{UU}(x_\pi,x_N,\vect{p}_T)$ for $0 \leqslant g_2 \leqslant 0.09\,$GeV, where for each $g_2$ the normalization constant $\mathcal{N}$ is chosen so that this result equals $P(x_F,\vect{p}_T;m_{\mu \mu}) / |\vect{p}_T|$ at $|\vect{p}_T| = 0.125\,$GeV, which represents the lowest $|\vect{p}_T|$ value in the E-615 data set~\cite{Stirling:1993gc}. Since Eq.~\eqref{eq:fuu} only describes the $W$-term we restrict $|\vect{p}_T| \leqslant 0.2\,m_{\mu \mu}$ following the finding of Ref.~\cite{Scimemi:2017etj}.  To study the ``prescription dependence'' of the TMD evolution, we also present evolved TMD results using the $\zeta$-prescription~\cite{Scimemi:2017etj} as the  dashed lines in Fig.~\ref{fig:p}, where we have taken $g_2 = 0$. As made clear from Fig.~\ref{fig:p} the two evolution prescriptions give similar results, and our results for the fitting function $P$ at $x_F = 0, 0.25$ are in good agreement with the E-615 data. For $x_F=0.5$ we find a discrepancy with data of around $30\%$, however for each $x_F$ our results favor a small value for $g_2$ as suggested in Ref.~\cite{Scimemi:2017etj}. In general, agreement with data could be improved by increasing the initial scale of the DSE calculations, which is an indication that higher Fock-states could play an important role.

Using the DCSB-improved truncation to QCD's DSEs we have determined the pion's minimal Fock-state LFWFs from the solution to the BSE, and from these LFWFs the pion's leading-twist time-reversal even TMD. The pion, as the Goldstone boson associated with DCSB in QCD, provides the ideal environment to study the impact of DCSB on hadron structure. We find that DCSB effects produce broad unimodal LFWFs and TMD when viewed as a function of $x$, for small $\vect{k}_T^2 \lesssim \L_{\rm QCD}^2$. In this regime the $\vect{k}_T^2$ dependence of the pion's LFWFs and TMD, for a given $x$, is well described by a Gaussian, however the $x$ and $\vect{k}_T^2$ does not factorize. These DCSB driven effects diminish slowly as $\vect{k}_T^2$ becomes large, where for $\vect{k}_T^2 \gtrsim 10\,$GeV$^2$ the LFWFs scale as $\psi_0 \propto x(1-x)/\vect{k}_T^2$  and  $\psi_1 \propto x(1-x)/\vect{k}_T^4$ and therefore agree with the power-law behavior predicted by perturbative QCD. We therefore make the prediction that in this regime both LFWFs have the same $x$ dependence. For large $\vect{k}_T^2$ the TMD exhibits two scaling regimes, first scaling like $f_\pi^q(x,\vect{k}_T^2) \propto 1/\vect{k}_T^6$ in the domain dominated by $\psi_1(x,\vect{k}_T^2)$ and then for $\vect{k}_T^2 \gtrsim 100\,$GeV$^2$ the asymptotic regime is reached where the TMD behaves as $f_\pi^q(x,\vect{k}_T^2) \propto x^2(1-x)^2/\vect{k}_T^4$.  By combining our predictions for the pion's TMD with empirical results for the nucleon's unpolarized TMD we made a comparison with data from the E-615 pion-induced Drell-Yan experiment, finding good agree for $0 < x_F < 0.25$. These results illustrated how a momentum tomography for the pion can shed like on hadron structure effects driven by DCSB and also help expose the transition from the non-perturbative to perturbative regimes in QCD.

%===============================================================================
%===============================================================================
\begin{acknowledgments}
C.S. thanks Cédric Mezrag for numerous helpful conversations, and Alexey Vladimirov for generous assistance with arTeMiDe. 
This work was supported by the U.S. Department of Energy, Office of Science, Office of Nuclear Physics, contract no. DE-AC02-06CH11357; and the Laboratory Directed Research and Development (LDRD) funding from Argonne National Laboratory, project no. 2016-098-N0 and project no. 2017-058-N0.
\end{acknowledgments}

%\bibliography{tmd}
%\bibliography{bibtexfile,bibtexfile_cloet,bibtexfile_books,TMD}


\end{document}

Our DSE results for the valence $u$ quark distribution in the $\pi^+$ at a scale of $Q^2 = 16\,$GeV$^2$ are given as the solid line in Fig.~\ref{fig:pdf1}. The Fermilab E-615 pion-induced Drell-Yan data~\cite{Conway:1989fs} are also given, toegther with a NLO re-analysis which includes soft-gluon resummation~\cite{Aicher:2010cb}. For moderate $0.2 \lesssim x \lesssim 0.6$ our result agrees well with the experimental data and the re-analysis from Ref.~\cite{Aicher:2010cb}. However, as $x$ increases our PDF deviates from the E-615 data but continues to agree reasonably well with Ref.~\cite{Aicher:2010cb}. Therefore our DSE result, and the re-analysis from Ref.~\cite{Aicher:2010cb}, both agree with the perturbative QCD prediction that the pion's valence PDF should behave as $f_\pi(x) \stackrel{x\to 1}{\propto} (1-x)^{\beta}$ with $\beta \geqslant 2$. In our calculation this condition is satisfied because at hadronic scale $f_\pi(x) \propto (1-x)^2$ as $x\to 1$, and the effect of the DGLAP evolution is to shift the exponent to larger values.

The unpolarized-unpolarized Drell-Yan differential cross section is generally~\cite{Arnold:2008kf}
\begin{align}
\frac{d \sigma}{d^4 q d \Omega}&=\frac{\alpha^2_{\textrm em}}{F q^2}\{(1+{\rm cos}^2\theta)F_{UU}^1+(1-{\rm cos}^2\theta)F_{UU}^2+  \nonumber \\
& +{\rm sin}2\theta {\rm cos}\phi F_{UU}^{{\rm cos}\phi}+ \sin^2\theta \cos 2 \phi F_{UU}^{{\rm cos}2\phi}\}.
\end{align}
After integrating out the angular dependence, at leading twist only the structure function $F^1_{UU}$ survives.